# Comparison of some properties of star forming galaxies and active galactic nuclei (AGNs) between two BOSS galaxy samples of SDSS DR9

Xin-Fa Deng

School of Science, Nanchang University, Jiangxi, China, 330031

Abstract Using the LOWZ and CMASS samples of the ninth data release (DR9) from the SDSS-III Baryon Oscillation Spectroscopic Survey (BOSS), I investigate properties of star forming galaxies and AGNs. The CMASS sample seriously suffers from the radial selection effect, even within the redshift $0.44 \leq z \leq 0.6$, which will lead to statistical conclusions in the CMASS sample being less likely robust. In the LOWZ sample, the fraction of star-forming galaxies is nearly constant from the lowest density regime to the densest regime; the AGN fraction is also insensitive to the local environment. In addition, I note that in the LOWZ sample, stellar mass and stellar velocity dispersion distributions of star forming galaxies and AGNs are nearly same.

## 1. INTRODUCTION

In the past, some issues of active galactic nuclei (AGNs) have been controversial. For example, the AGN fraction obtained in some works was fairly different, from a few percent to $\approx 40\%$ (Dressler et al. 1985; Huchra & Burg 1992; Ho et al. 1997; Carter et al. 2001; Ivezic′ et al. 2002; Miller et al. 2003; Deng et al. 2012a, 2012c). This can be traced to two facts. First, these authors used different galaxy samples. Second, they applied different AGN classification techniques. Many works also focused on the environmental dependence of the AGN fraction, which can provide tests of models of AGN activity and formation. In this aspect, typical works included ones of Carter et al. (2001) and Miller et al. (2003), who reported that the AGN fraction depends very little on the environment. But some authors argued that the AGN fraction should decrease with increasing density (e.g., Dressler et al. 1985; Kauffmann et al. 2004; Popesso & Biviano 2006; von der Linden et al. 2010; Deng et al. 2012c). Different environmental dependence of the AGN fraction actually means different physical mechanisms of AGN activity and formation.

Galaxy samples often were divided into two opposite families, such as luminous and faint, red and blue, early-type and late-type. One often performed comparative studies between two opposite families. For example, Deng et al. (2009a) and Deng (2010) compared some properties of early-type galaxies with those of late-type galaxies. Deng et al. (2011b) explored the environmental dependence of star formation rate (SFR), specific star formation rate (SSFR) and stellar mass for blue and red galaxies. Here, as a comparison, I also investigate some properties of star forming galaxies.

The SDSS-III (Eisenstein et al. 2011) Baryon Oscillation Spectroscopic Survey (BOSS) is a valuable project, which will carry out a redshift survey of 1.5 million luminous red galaxies (LRGs) at $0.15 < z < 0.8$ over 10, 000 square degrees and 160,000 quasi-stellar objects (QSOs) at 2.15 < z < 3.5 over 8,000 square degree. The public data release of BOSS spectra data includes a number of physical galaxy parameters derived by some authors. Using principal component analysis (PCA), Chen et al. (2012) estimated stellar masses of $\approx$ 290, 000 BOSS galaxies with stellar masses of >$10^{11}$ $M_\odot$ and a redshift range of 0.4 < z < 0.7. By fitting model spectral energy distributions to u, g, r, i, z magnitudes, Maraston et al. (2012) calculated stellar masses for $\approx$ 400, 000 BOSS galaxies at redshift $\approx$ 0.2-0.7. Thomas et al. (2013) measured stellar velocity dispersion and presented the BPT classification (Baldwin et al. 1981) of BOSS galaxies. These data sets will provide source material for the study of many issues of galaxies.

My paper is organized as follows. In section 2, I describe the data used. The density estimator and results are presented in section 3 and section 4. In section 5, I compare some results of this work with previous ones. My main results and conclusions are summarized in section 6.

In calculating the distance I used a cosmological model with a matter density $\Omega_0 = 0.3$, cosmological constant $\Omega_\Lambda = 0.7$, Hubble's constant $H_0 = 70 \mathrm{km} \cdot \mathrm{s}^{-1} \cdot \mathrm{Mpc}^{-1}$.

2. **Data**

The ninth data release (DR9) (Ahn et al. 2012) of the SDSS is the first public release of spectroscopic data from the SDSS-III Baryon Oscillation Spectroscopic Survey (BOSS), which includes 535,995 new galaxy spectra (median z $\approx$ 0.52), 102,100 new quasar spectra (median z $\approx$ 2.32), and 90,897 new stellar spectra, along with the data presented in previous data releases.

The BOSS galaxy sample is divided into two principal samples at z $\approx$ 0.4: "LOWZ" and "CMASS". The LOWZ sample is a low redshift sample, with a median redshift of z = 0.3, containing about a quarter of all galaxies in BOSS. This sample is a simple extension of the SDSS-I and –II Luminous Red Galaxy (LRG) sample (Eisenstein et al. 2001) to lower luminosities and allows for a comparison with the SDSS I & II samples. The space density of LOWZ galaxies is about 2.5times that of the SDSS-I/II LRG sample. The CMASS sample, containing three times more galaxies than LOWZ, is designed to select galaxies above z $\approx$ 0.4, and is a nearly complete sample of massive galaxies above the magnitude limit of the survey. The LOWZ sample mostly contains galaxies at 0.15 < z < 0.43, while the CMASS sample mostly contains galaxies at 0.43 < z < 0.7.

In this work, the data was downloaded from the Catalog Archive Server of SDSS Data Release 9 (Ahn et al. 2012) by the SDSS SQL Search (http://www.sdss3.org/dr9/). I extract 85360 LOWZ galaxies with the redshift 0.15 ≤ z ≤ 0.43 (with SDSS flag: BOSS_TARGET1&1>0) and 301190 CMASS galaxies with the redshift 0.43 ≤ z ≤ 0.7 (with SDSS flag: BOSS_TARGET1&128>0).

Maraston et al. (2012) employed the two template fittings (passive and star-forming ) and the two adopted Initial Mass Functions (IMFs) (Salpeter and Kroupa ) , and also considered the mass lost via stellar evolution. In this work, I use the data of best-fit stellar mass [in log $M_{sun}$] obtained with the star-forming template and the Kroupa IMF (http://data.sdss3.org/sas/dr9/boss/spectro/redux/galaxy/). Stellar mass of few galaxies is 0 by definition. I remove these galaxies from samples. Finally, the total galaxy number of the LOWZ sample is reduced to 85295, while the total galaxy number of the CMASS sample is reduced to 296501.

Thomas et al. (2013) performed a spectroscopic analysis of galaxy spectra from the SDSS-III Baryon Oscillation Spectroscopic Survey (BOSS). This data set is also released in the Web: http://data.sdss3.org/sas/dr9/boss/spectro/redux/galaxy/. BPT classification first introduced by Baldwin et al. (1981) was widely used in studies of SDSS galaxies, which has become standard practice to classify objects according to their position on the so-called BPT diagrams. Thomas et al. (2013) adopted the empirical separation between star forming galaxies and AGN defined by Kauffmann et al. (2003). Star forming galaxies are the galaxies that lie below this line in BPT diagrams. The AGN population consists of the galaxies above the theoretical extreme star burst line developed by Kewley et al. (2001). Composite galaxies are the objects that are between these two separating lines. This area is populated by galaxies with a composite of star burst and AGN spectra. Thomas et al. (2013) further used the dividing line defined by Schawinski et al. (2007) to distinguish between LINER and Seyfert emission based on SDSS galaxy classifications obtained through the [NII]/H$\alpha$ ratio. In this data set, all analyses are restricted to those spectra where the full set of key diagnostic emission lines H$\beta$, [OIII], H$\alpha$, and [NII] can be detected with an AoN above 1.5 (requiring the amplitude-over-noise ratio of all four lines to be larger than 1.5) to allow for a proper analysis of the emission line characteristics through emission line ratio diagnostic diagrams.

## 3. Density estimator

In this work, following Deng (2010), the three-dimensional local galaxy density LD (Galaxies Mpc$^{-3}$) is computed in a comoving sphere with a radius of the distance to the 5th nearest galaxy for each galaxy. To explore environmental dependence of BOSS galaxy properties, like Deng et al. (2008) did, I arrange galaxies in a density order from the smallest to the largest, select approximately 5% of the galaxies, construct two subsamples at both extremes of density according to the density, and compare BOSS galaxy properties in the lowest density regime with those in the densest regime.

## 4. Comparison of some properties of star forming galaxies and AGNs between two BOSS galaxy samples of SDSS DR9

Table 1 shows galaxy number and fraction of different classes in the LOWZ and CMASS samples and different subsamples. As seen from this table, the fraction of star-forming galaxies in the LOWZ sample is nearly constant from the lowest density regime to the densest regime.

In the CMASS sample with the redshift 0.43 ≤ z ≤ 0.7, the fraction of star-forming galaxies dramatically increases with increasing density. This is likely due to the radial selection effect in the CMASS sample. Dawson et al. (2013) argued that the BOSS galaxies are selected to have approximately uniform comoving number density of $\bar{n} = 3 \times 10^{-4}$ h$^3$ Mpc$^{-3}$ out to a redshift z = 0.6, then monotonically decreasing to zero density at z $\approx$ 0.8. Fig.2 of Anderson et al. (2012) showed that the number-density of CMASS galaxies dramatically drops with increasing redshift at redshift z>0.6. Thus, CMASS galaxies with low density exist preferentially in the redshift range

z>0.6, which leads to CMASS subsample at the lowest density regions containing lower fraction of star-forming galaxies (Thomas et al. 2013).

Thomas et al. (2013) argued that the fraction of star forming galaxies decreases and the fraction of AGN increases with increasing redshift, mostly owing to selection effects. Fig.1 shows the fraction of BLANK, Star Forming, Composite and AGN(Seyfert+ LINER+ Seyfert/ LINER) classes as a function of redshift for the LOWZ and CMASS samples. Fig.2 further demonstrates the fraction of different AGNs (Seyfert, LINER and Seyfert/ LINER) as a function of redshift for the LOWZ and CMASS samples. In the LOWZ sample, selection effects are not very serious: the fraction of star forming galaxies is weakly redshift dependent and the fraction of AGN slightly increases with increasing redshift. In the CMASS sample, the fraction of AGN decreases with increasing redshift and the fraction of star forming galaxies slightly increases with redshift. It is noteworthy that at $z \approx 0.6$, the fraction of Star Forming, Composite and AGN classes dramatically drops to zero, while the fraction of BLANK class reaches to 1. This shows that selection effects in the CMASS sample are fairly serious. Maraston et al. (2012) also claimed that BOSS is a mass-uniform sample over the redshift range 0.2 to 0.6. Thus, I construct a CMASS sample with the redshift $0.44 \leq z \leq 0.6$, which contains 224435 galaxies. This CMASS sample should be a relatively uniform sample, in which the radial selection effect is less important. In the following analyses, I only use this CMASS sample. As seen in table 1, in the CMASS sample with the redshift $0.44 \leq z \leq 0.6$, the fraction of star-forming galaxies still increases with increasing density.

Table1 demonstrates that the fraction of each class in two BOSS galaxy samples is fairly different, which shows that the fraction of each class is seriously influenced by properties of BOSS galaxy samples.

The study of the environmental dependence of the AGN fraction has long been an important issue, which can provide tests of models of AGN activity and formation. Miller et al. (2003) argued that independence on local galaxy density of the AGN fraction may mean that the AGN population is primarily tracing only the bulge component of galaxies. Miller et al. (2003) also claimed that if there is a strong relationship between the presence of an AGN and the ability of a galaxy to form stars, the AGN fraction should decrease with increasing density, analogous to the SFR(star-formation rate) -density relation (e.g., Balogh et al. 1998; Hashimoto et al. 1998; Lewis et al. 2002; Gómez et al. 2003; Tanaka et al. 2004; Patel et al. 2009). If galaxy-galaxy collisions fuel the AGN activity by driving gas into the cores of galaxies and thus onto the black hole (BH; Gunn 1979; Shlosman et al. 1990), the fraction of galaxies with an AGN should increase with increasing local density.

As seen from table 1, the AGN (Seyfert+ LINER+ Seyfert/ LINER) fraction of two subsamples at both extremes of density in the LOWZ sample is: 59.8% for subsample at low density and 60.1% for subsample at high density, which shows that the AGN fraction is insensitive to the local environment. However, in the CMASS sample with the redshift $0.44 \leq z \leq 0.6$, the AGN fraction increases considerably with increasing density: 37.4% for subsample at low density and 49.4% for subsample at high density. This is still due to the radial selection effect. As seen from Fig.1, at $z \approx 0.6$, the AGN fraction dramatically drops to zero. Fig.2 of Anderson et al. (2012) showed that the number-density of CMASS galaxies reaches to the peak value at $z \approx 0.53$, then decreases substantially with increasing redshift. Thus, most CMASS galaxies with low density are likely located at $z \approx 0.6$, which leads to CMASS subsample at the lowest density regions containing lower AGN fraction.

It is noteworthy that the fraction of BLANK class in the CMASS sample is fairly high, which means that classification of too many galaxies could not be made. Such a drawback will lead to statistical conclusions in the CMASS sample being less likely robust. CMASS galaxies mostly are located at z>0.45. Thomas et al. (2013) indicated that the analysis of AGN fractions at high redshifts requires alternative methods for the identification of AGN, which is a future work.

Fig.3 shows stellar mass distribution of star forming galaxies and AGNs for the LOWZ (a) and CMASS (b) samples. In the LOWZ sample, stellar mass distribution of star forming galaxies and AGNs is nearly same. In the CMASS sample, a weak trend can be observed: AGNs are preferentially more massive. But this is likely due to the radial selection effect.

Fig.4 shows stellar velocity dispersion distribution of star forming galaxies and AGNs for the LOWZ (a) and CMASS (b) samples. Only in the CMASS sample, an apparent trend exists: AGNs have preferentially larger stellar velocity dispersion. Thomas et al. (2013) argued that the typical velocity dispersion of a BOSS galaxy is $\approx 240$ km s$^{-1}$. Indeed, the peak value of stellar velocity dispersion distribution in two samples is at $\approx 240$ km s$^{-1}$.

Table 1: Galaxy number and fraction of different classes in the LOWZ and CMASS samples and different subsamples.
5

| Sample | All N (%) | BLANK N (%) | Star Forming N (%) | Composite N (%) | Seyfert N (%) | LINER N (%) | Seyfert/ LINER N (%) |
|---|---|---|---|---|---|---|---|
| LOWZ sample (0.15 ≤ z ≤ 0.43) | 85295 (100.0) | 9364(11.0) | 10511(12.3) | 12019(14.1) | 21201(24.9) | 26670(31.3) | 5530(6.4) |
| LOWZ subsample (0.15 ≤ z ≤ 0.43) at the lowest density regions (LD=5.18×10$^{-8}$--1.71×10$^{-5}$ galaxies Mpc$^{-3}$) | 4265 (100.0) | 440(10.3) | 574(13.5) | 700(16.4) | 988(23.2) | 1324(31.0) | 239(5.6) |
| LOWZ subsample (0.15 ≤ z ≤ 0.43) at the densest density regions (LD=1.11×10$^{-3}$—3.91×10$^{-2}$ galaxies Mpc$^{-3}$) | 4265 (100.0) | 464(10.9) | 580(13.6) | 655(15.4) | 942(22.1) | 1355(31.7) | 269(6.3) |
| CMASS sample (0.43 ≤ z ≤ 0.70) | 296501 (100.0) | 130963 (44.2) | 42052 (14.2) | 19474 (6.6) | 42130 (14.2) | 44705 (15.1) | 17177 (5.7) |
| CMASS subsample (0.43 ≤ z ≤ 0.70) at the lowest density regions (LD=1.44×10$^{-6}$—2.39×10$^{-5}$ galaxies Mpc$^{-3}$) | 14825 (100.0) | 12296(82.9) | 611(4.1) | 334(2.3) | 666(4.5) | 689(4.6) | 229(1.6) |
| CMASS subsample (0.43 ≤ z ≤ 0.70) at the densest density regions (LD=1.75×10$^{-3}$--1.16×10$^{-1}$ galaxies Mpc$^{-3}$) | 14825 (100.0) | 3904(26.3) | 2614(17.6) | 1199(8.1) | 2788(18.8) | 3064(20.7) | 1256(8.5) |
| CMASS sample (0.44 ≤ z ≤ 0.60) | 224435 (100.0) | 62534(27.9) | 41296(18.4) | 18991(8.5) | 41019(18.3) | 43706(19.5) | 16889(7.4) |
| CMASS subsample (0.44 ≤ z ≤ 0.60) at the lowest density regions (LD=3.52×10$^{-6}$—3.83×10$^{-5}$ galaxies Mpc$^{-3}$) | 11222 (100.0) | 4425(39.4) | 1746(15.6) | 848(7.6) | 1712(15.3) | 1834(16.3) | 657(5.8) |
| CMASS subsample (0.44 ≤ z ≤ 0.60) | 11222 (100.0) | 2686(23.9) | 2044(18.2) | 951(8.5) | 2165(19.3) | 2390(21.3) | 986(8.8) |

| | | | | | | |
|---|---|---|---|---|---|---|
| at the densest density regions (LD=2.05×10$^{-3}$--1.16×10$^{-1}$ galaxies Mpc$^{-3}$) | | | | | | |

Notes: 1. BPT='BLANK' means no classification could be made, usually because one of the lines involved has a 'NaN' or 'Inf' value in the catalogue. 2. Seyfert/LINER are those cases where the line ratios happen to sit exactly on the dividing line.

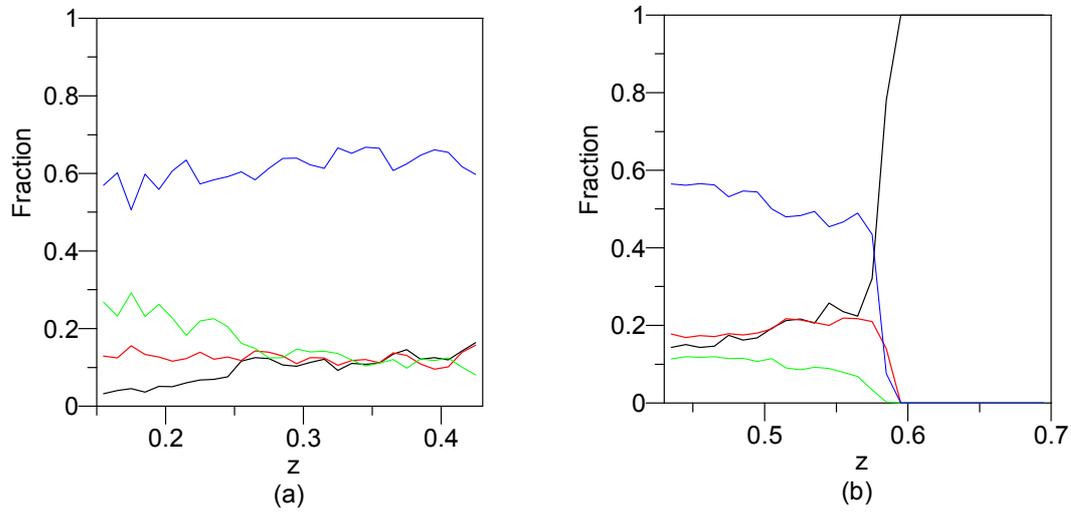

Fig.1 Fraction of different classes as a function of redshift for the LOWZ (a) and CMASS (b) samples: black, red, green and blue lines represent BLANK, Star Forming, Composite and AGN(Seyfert+ LINER+ Seyfert/ LINER), respectively.

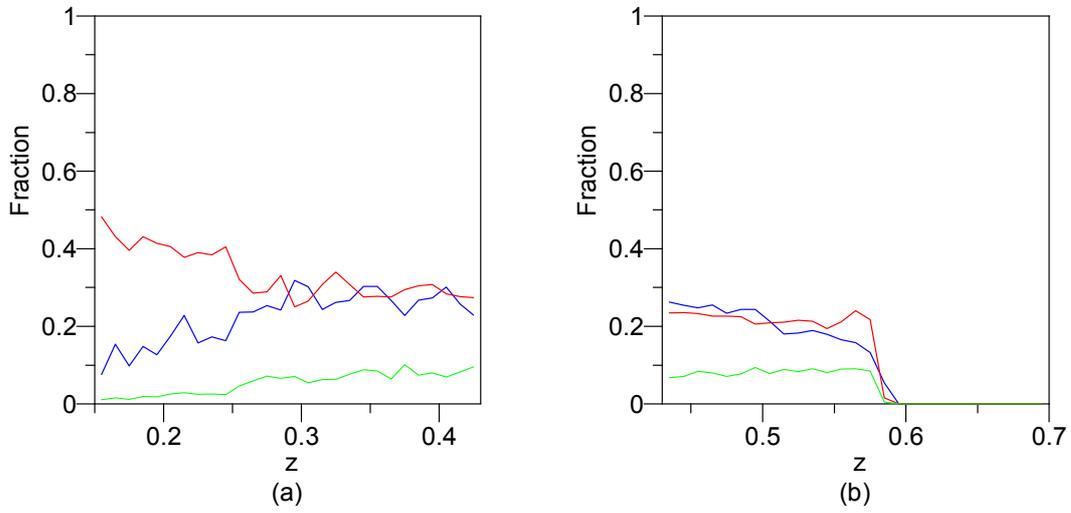

Fig.2 Fraction of different AGNs as a function of redshift for the LOWZ (a) and CMASS (b) samples: blue, red, green lines represent Seyfert, LINER and Seyfert/ LINER, respectively.

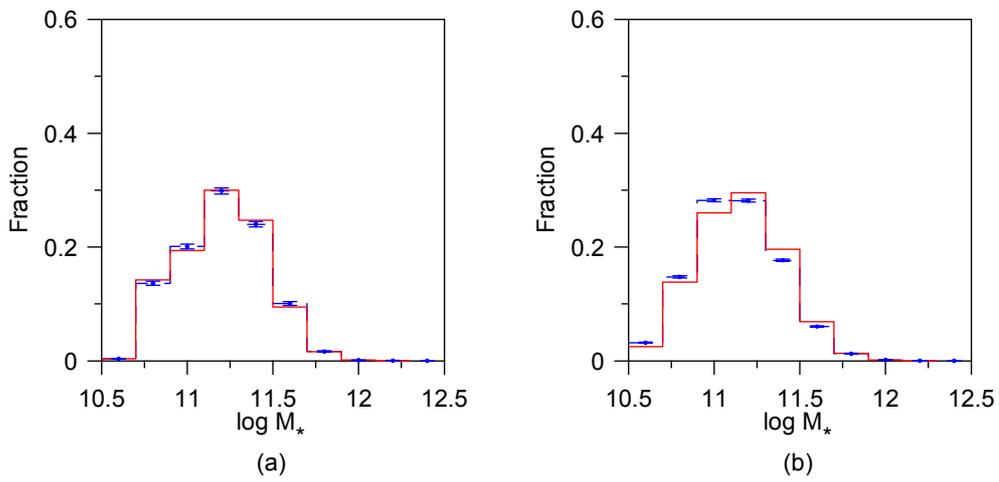

Fig.3 Stellar mass distribution of star forming galaxies and AGNs for the LOWZ (a) and CMASS (b) samples: red solid line for AGNs, blue dashed line for star forming galaxies. The error bars of blue lines are 1 $\sigma$ Poissonian errors. Error-bars of red lines are omitted for clarity.

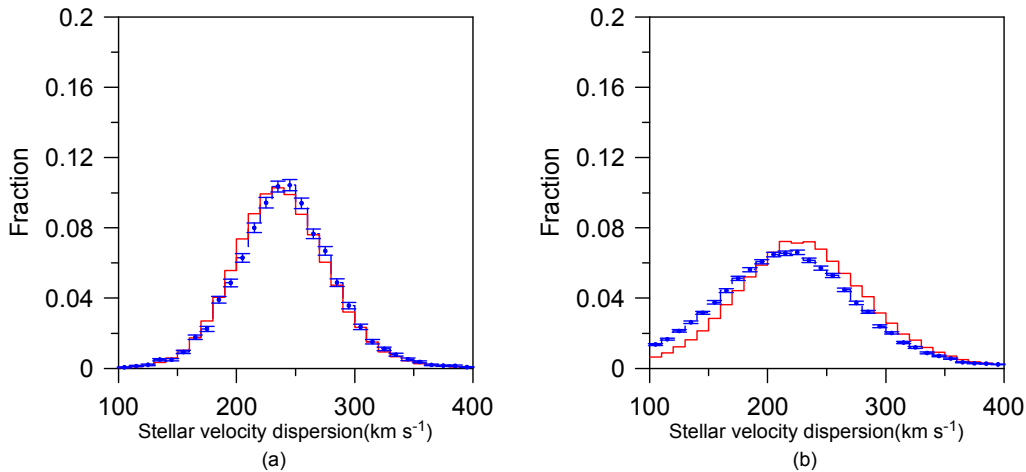

Fig.4 Stellar velocity dispersion distribution of star forming galaxies and AGNs for the LOWZ (a) and CMASS (b) samples: red solid line for AGNs, blue dashed line for star forming galaxies. The error bars of blue lines are 1 $\sigma$ Poissonian errors. Error-bars of red lines are omitted for clarity.

## 5. Comparison with previous results

Deng et al. (2012a) reported that in the two volume-limited Main galaxy (Strauss et al. 2002) samples of the Sloan Digital Sky Survey Data Release 8 (the SDSS DR8) (Aihara et al. 2011), the fraction of star-forming galaxies decreases with density, which is in good agreement with previous results (Carter et al. 2001; Miller et al. 2003; Deng et al. 2009b, 2011a). Such a trend means that high density environments tend to suppress star formation(e.g., Balogh et al. 1998; Hashimoto et al. 1998; Lewis et al. 2002; Go´mez et al. 2003; Tanaka et al. 2004; Patel et al. 2009), which is suggested by many possible mechanisms(Gunn & Gott 1972; Byrd & Valtonen 1990; Zabludoff et al. 1996; Zabludoff & Mulchaey 1998; Moore et al. 1999; Quilis et al. 2000). However, Deng et al. (2012b) found that the environmental dependence of star formation rate (SFR) and specific star formation rate (SSFR) in the LRG sample is fairly weak. LRGs are a group of galaxies that are likely to be luminous, red and early-types. Deng et al. (2012b) argued that weak environmental dependence of SFR and SSFR in the LRG sample can be traced to two facts. First, galaxy color and morphology are a pair of galaxy properties most predictive of local environments. Second, strong environmental dependence of galaxy properties for red galaxies mainly is due to the one in the red late-type sample (Deng et al. 2011b). As indicated as the above-mentioned text, the LOWZ sample only is a simple extension of the SDSS-I and –II Luminous Red Galaxy (LRG) sample. Thus, it is not surprising that there nearly is no correlation between the fraction of star-forming galaxies and the local density in the LOWZ sample.

As seen in table 1, the AGN(Seyfert, LINER and Seyfert/ LINER) fraction in the LOWZ sample is $\approx 62.6\%$ and this fraction in the CMASS sample with the redshift $0.44 \leq z \leq 0.6$ is $\approx 45.2\%$, which is much larger than results obtained in previous works. Dressler et al. (1985) and Huchra & Burg (1992) reported that only a few percent of galaxies possessed an AGN. Ivezic′ et al. (2002) also claimed that only 5% of SDSS galaxies had an AGN. Miller et al. (2003) showed that the AGN fraction in the early data release of the SDSS (Stoughton et al. 2002) is about 20%, which is closer to that found in the 15R-North galaxy redshift survey by Carter et al. (2001). In two volume-limited Main galaxy samples of the SDSS DR6 (Adelman-McCarthy et al. 2008) above and below the value of $M_r^*$, Deng et al. (2012c) demonstrated that the AGN fraction in the luminous volume-limited sample is $\approx 9.8\%$ and the AGN fraction in the faint volume-limited sample is $\approx 4.3\%$. Deng et al. (2012a) argued the difference of the AGN fraction obtained by different authors likely is due to these authors using different AGN classification techniques. However, in works of Deng et al. (2012a) and Deng et al. (2012c), two volume-limited Main galaxy samples above and below the value of $M_r^*$ contain fairly different AGN fraction, even if the same AGN classification techniques were used in two samples. Deng et al. (2012a) and Deng et al. (2012c) argued that the AGN fraction also is seriously influenced by properties of galaxy samples. Deng et al. (2012a) used galaxy physical parameters derived by the MPA-JHU group (http://www.mpa-garching.mpg.de/SDSS/DR7/), such as BPT classification, stellar mass, nebular oxygen abundance, star formation rates (SFRs) and the specific SFR (SSFR). BPT classification in this Web is based on the methodology of Brinchmann et al. (2004), which is completely the same as the methodology of Thomas et al. (2013) adopted in this work. But the AGN fraction in two BOSS galaxy samples is much larger than that found in two volume-limited Main galaxy samples by Deng et al. (2012a). This further shows that the AGN fraction indeed is correlated with some properties of galaxy samples.

Kauffmann et al. (2003) argued that AGNs of all luminosities of the [OIII] $\lambda$ 5007 emission line reside almost exclusively in massive galaxies. Deng et al. (2012a) concluded that galaxies hosting AGNs are preferentially more massive. For a long time, the supermassive black holes were hypothesized as the power source of AGNs (Salpeter 1964; Lynden-Bell 1969). Heckman et al. (2004) reported that most present-day accretion occurs onto black holes that reside in moderately massive galaxies ($M_* \approx 10^{10} - 10^{11.5} M_\odot$). However, I calculate the fraction of moderately massive galaxies in different galaxy samples: 88.75% in the LOWZ sample, 90.88% in the CMASS sample with the redshift 0.44 <z<0.6, 32.86% in the faint volume-limited sample of Deng et al. (2012a) and 98.49% in the luminous volume-limited sample of Deng et al. (2012a), which does not support this standpoint. Deng et al. (2012c) also argued that the presence of AGNs is not correlated with the stellar mass. They demonstrated that about 70.82% of luminous hosts with an AGN are high mass galaxies, while this fraction in faint hosts with an AGN is only 3.73%.

Carter et al. (2001) and Deng et al. (2012c) demonstrated that the AGN fraction increases steeply with luminosity. BOSS galaxies are a group of fairly luminous galaxies, compared with previous galaxy samples. High fraction of AGN in BOSS galaxy samples like is due to the correlation between the presence of AGNs and luminosity.

Numerous authors showed that the AGN fraction nearly is not correlated with the local environment of AGN host galaxies (e.g., Monaco et al. 1994; Coziol et al. 1998; Shimada et al. 2000; Carter et al. 2001; Schmitt 2001; Miller et al. 2003; Pasquali et al. 2009). However, there have been a number of dissenting papers. Dressler et al. (1985) and Popesso & Biviano (2006) reported a lower fraction of AGNs in clusters than in the field. Kauffmann et al. (2004) argued that galaxies in dense environments are less likely to host a powerful optical AGN (L[OIII] > $10^7 L_\odot$). von der Linden et al.(2010) claimed that the overall fraction of strong AGNs decreases towards the cluster center. Deng et al. (2012c) showed the fraction of AGNs declines substantially with increasing local density in the luminous volume-limited sample, but in the faint volume-limited sample this change is very weak. Deng et al. (2012a) argued that such a controversial conclusion likely also is due to these authors using different AGN classification techniques and galaxy samples. For example, in the two volume-limited Main galaxy samples of the SDSS DR8, Deng et al. (2012a) found that AGN fraction is nearly insensitive to the local environment, which is not consistent with the conclusion of Deng et al. (2012c). Deng et al. (2012c) used the empirical demarcation line between star-forming galaxies and AGNs developed by Kauffmann et al. (2003). Thus, AGN samples of Deng et al. (2012c) actually contain composite galaxies (**C**) and AGN populations. Deng et al. (2012a) indicated that if composite galaxies (**C**) are classified as AGN like Deng et al. (2012c) did, the fraction of AGNs in the luminous volume-limited sample declines with increasing local density.

Kauffmann et al. (2004) argued that the fraction of strong AGNs in massive galaxies *decreases* as a function of density, while the fraction of *low-luminosity* AGNs depends very little on local density. Miller et al. (2003) reported that the AGN fraction shows little dependence on local density. Kauffmann et al. (2004) claimed that it is likely due to their sample containing a substantial number of weak AGNs. Kauffmann et al. (2004) indicated that AGNs with L[OIII] > $10^7 L_\odot$ are almost all type 2 Seyfert galaxies, while *low-luminosity* AGNs are LINERs. However, table 1 demonstrates that in the LOWZ sample, both the fraction of Seyfert and LINER classes are nearly independent of the local environment. As indicated as Deng et al. (2012c), this shows that the classification of AGNs does not essentially decide whether the AGN fraction depends on the environment.

## 6. Summary

Using the LOWZ and CMASS samples of the ninth data release (DR9) from the SDSS-III Baryon Oscillation Spectroscopic Survey (BOSS), I investigate properties of star forming galaxies and AGNs. The main results can be summarized as follows:

1) The CMASS sample seriously suffers from the radial selection effect, even within the redshift $0.44 \leq z \leq 0.6$, which will lead to statistical conclusions in the CMASS sample being less likely robust. Thomas et al. (2013) indicated that the analysis of AGN fractions at high redshifts requires alternative methods for the identification of AGN. Improvement of all these drawbacks in CMASS sample should be a focus of future works.

2) The fraction of each class in two BOSS galaxy samples is fairly different, which shows that the fraction of each class is seriously influenced by properties of BOSS galaxy samples.

3) In the LOWZ sample, the fraction of star-forming galaxies is nearly constant from the lowest density regime to the densest regime, which is consistent with results of LRGs.

4) The AGN(Seyfert, LINER and Seyfert/ LINER) fraction in the LOWZ sample is $\approx 62.6\%$ and this fraction in the CMASS sample with the redshift $0.44 \leq z \leq 0.6$ is $\approx 45.2\%$, which is much larger than results obtained in previous works.

5) In the LOWZ sample, the AGN fraction is insensitive to the local environment.

6) In the LOWZ sample, stellar mass and stellar velocity dispersion distributions of star forming galaxies and AGNs are nearly same.

**Acknowledgements**

I thank the anonymous referee for many useful comments and suggestions. This study was supported by the National Natural Science Foundation of China (NSFC, Grant 11263005).

Funding for SDSS-III has been provided by the Alfred P. Sloan Foundation, the Participating Institutions, the National Science Foundation, and the U.S. Department of Energy. The SDSS-III web site is http://www.sdss3.org/.

SDSS-III is managed by the Astrophysical Research Consortium for the Participating Institutions of the SDSS-III Collaboration including the University of Arizona, the Brazilian Participation Group, Brookhaven National Laboratory, University of Cambridge, University of Florida, the French Participation Group, the German Participation Group, the Instituto de Astrofisica de Canarias, the Michigan State/Notre Dame/JINA Participation Group, Johns Hopkins University, Lawrence Berkeley National Laboratory, Max Planck Institute for Astrophysics, New Mexico State University, New York University, Ohio State University, Pennsylvania State University, University of Portsmouth, Princeton University, the Spanish Participation Group, University of Tokyo, University of Utah, Vanderbilt University, University of Virginia, University of Washington, and Yale University.

**References**

Ahn C.P., Alexandroff R., Allende Prieto C. et al, 2012, ApJS, 203, 21
Adelman-McCarthy J.K., Agüeros M. A., Allam S.S. et al., 2008, ApJS, 175, 297
Aihara H., Prieto C. A., An D. et al., 2011, ApJS, 193, 29
Anderson L., Aubourg E., Bailey S. et al, 2012, MNRAS, 427, 3435
Baldwin J. A., Phillips M. M., Terlevich R., 1981, PASP, 93, 5 (BPT)
Balogh M.L., Schade D., Morris S.L. et al., 1998, ApJ, 504, L75
Brinchmann J., Charlot S., White S. D. M. et al., 2004, MNRAS, 351, 1151
Byrd G., & Valtonen M., 1990, ApJ, 350, 89
Carter B.J., Fabricant D.G., Geller M.J. et al., 2001, ApJ, 559, 606
Chen Y.M., Kauffmann G., Tremonti C. A. et al., 2012, MNRAS, 421, 314
Coziol R., de Carvalho R. R., Capelato H. V., & Ribeiro A. L. B., 1998, ApJ, 506, 545
Dawson K.S., Schlegel D.J., Ahn C.P. et al, 2013, AJ, 145, 10
Deng X.F., He J.Z., Song J. et al., 2008, PASP, 120, 487
Deng X.F., He J.Z., Wen X.Q., 2009a, ApJ, 693, L71
Deng X.F., He J.Z., Chen Y.Q. et al., 2009b, ApJ, 706, 436
Deng X.F., 2010, ApJ, 721, 809
Deng X.F., Chen Y.Q., Jiang P. et al., 2011a, AN, 332, 706
Deng X.F., Chen Y.Q., Jiang P., 2011b, MNRAS, 417, 453
Deng X.F., Xin Y., Wu P. et al., 2012a, ApJ, 754, 82
Deng X.F., Yang B., Ding Y.P. et al., 2012b, AN, 333, 644
Deng X.F., Xin Y., Wu P. et al., 2012c, AN, 333, 767
Dressler A., Thompson I. B., & Shectman S. A., 1985, ApJ, 288, 481

Eisenstein D.J., Annis J., Gunn J. E. et al., 2001, AJ, 122, 2267

Eisenstein D. J., Weinberg D. H., Agol E. et al., 2011, AJ, 142, 72

Gómez P.L., Nichol R.C., Miller C.J. et al., 2003, ApJ, 584, 210

Gunn J. E., & Gott J. R. I., 1972, ApJ, 176, 1

Gunn J. E., 1979, in Active Galactic Nuclei, ed. C. Hazard & S. Mitton (Cambridge: Cambridge Univ. Press), 213

Hashimoto Y., Oemler A., Lin J.H. et al., 1998, ApJ, 499, 589

Heckman T. M., Kauffmann G., Brinchmann J. et al., 2004, ApJ, 613, 109

Ho L. C., Filippenko A. V., & Sargent W. L. W., 1997, ApJ, 487, 568

Huchra J., & Burg R., 1992, ApJ, 393, 90

Ivezić Z. et al., 2002, AJ, 124, 2364

Kauffmann G., Heckman T. M., Tremonti C. et al., 2003, MNRAS, 346, 1055

Kauffmann G., White S. D. M., Heckman T. M. et al., 2004, MNRAS, 353, 713

Kewley L. J., Dopita M. A., Sutherland R. S. et al., 2001, ApJ, 556, 121

Lewis I., Balogh M., Propris R.D. et al., 2002, MNRAS, 334, 673

Lynden-Bell D., 1969, MNRAS, 143, 167

Maraston C., Pforr J., Henriques B.M. et al., 2012, arXiv:1207.6114

Monaco P., Giuricin G., Mardirossian F., & Mezzetti M., 1994, ApJ, 436, 576

Miller C.J., Nichol R.C., Go′mez P. L. et al., 2003, ApJ, 597,142

Moore B., Lake G., Quinn T., & Stadel J., 1999, MNRAS, 304, 465

Pasquali A., van den Bosch F. C., Mo H. J. et al., 2009, MNRAS, 394, 38

Patel S.G., Holden B.P., Kelson D.D. et al., 2009, ApJ, 705, L67

Popesso P. & Biviano A. 2006, A&A, 460, L23

Quilis V., Moore B., & Bower R., 2000, Science, 288, 1617

Salpeter E., 1964, ApJ, 140, 796

Schawinski K., Thomas D., Sarzi M. et al., 2007, MNRAS, 382, 1415

Schmitt H. R., 2001, AJ, 122, 2243

Shimada M., Ohyama Y., Nishiura S. et al., 2000, AJ, 119, 2664

Shlosman I., Begelman M. C., & Frank J., 1990, Nature, 345, 679

Stoughton C., Lupton R. H., Bernardi M. et al., 2002, AJ, 123, 485

Strauss M. A., Weinberg D. H., Lupton R. H. et al., 2002, AJ, 124, 1810

Tanaka M., Goto T., Okamura S. et al.,2004, AJ, 128, 2677

Thomas D., Steele O., Maraston C. et al., 2013, MNRAS, 431, 1383

von der Linden A., Wild V., Kauffmann G. et al., 2010, MNRAS, 404, 1231

Zabludoff A. I., &Mulchaey J. S., 1998, ApJ, 496, 39

Zabludoff A. I., Zaritsky D., Lin H. et al., 1996, ApJ, 466, 104